\documentclass[10pt]{article}

\usepackage{cite}

\usepackage{graphicx}        
\usepackage{amssymb}
\usepackage{amsmath}
\usepackage{pdflscape}
\usepackage{tikz}
\usepackage[scriptsize]{subfigure}
\usepackage{amsmath}
\usepackage{soul}

\usepackage{float}

\usepackage[T1]{fontenc}
\usepackage{kpfonts,baskervald}

\usepackage[utf8]{inputenc}
\usepackage{authblk}
\usepackage{hyphenat}
\usepackage{cite}

\usepackage{placeins}

\usepackage{caption} 
\usepackage{dblfloatfix} 

\tikzset{cross/.style={cross out, draw=black, minimum size=2*(#1-\pgflinewidth), inner sep=0pt, outer sep=0pt},
cross/.default={2.5pt}}
\usepackage[margin=1in]{geometry}
\usepackage{parskip}
\usepackage{titlesec}

\usepackage{authblk} 

\usepackage{IEEEtrantools}

\usepackage{cancel}
\usepackage{multirow}

\DeclareGraphicsExtensions{.pdf,.png,.jpg,.eps} 
\usepackage{epstopdf}
\usepackage[acronym,shortcuts]{glossaries}

\makeglossaries
\usepackage{float}

\newacronym{GFVSC}{\textcolor{black}{GFM-VSC}}{Grid-forming voltage source converter}

\usepackage{multicol}

\begin{document}
%

\title{Active-power control strategies in grid-forming power converters to improve transient stability in power systems with 100\% converter-based generation}


 \author{R\'egulo E. \'Avila-Mart\'inez$^{1,2}$, Luis Rouco$^{1,2}$, Javier Renedo$^2$, Lukas Sigrist$^{1,2}$, Aurelio Garcia-Cerrada$^{1,2}$}
\date{%
    $^1$ Instituto de Investigación Tecnológica (IIT), Madrid, Spain. \\%
    $^2$ ETSI ICAI, Universidad Pontificia Comillas, Madrid, Spain. \\%
    \{regulo.avila, luis.rouco, lukas.sigrist, aurelio\}@iit.comillas.edu \\%
    javier.renedo@ieee.org \\ [2ex]%
    \today
}
 


\maketitle

\begin{abstract}
\acp{GFVSC} play a crucial role in the stability of power systems with large amounts of converter-based generation. Transient stability (angle stability under large disturbances) is  a critical limiting factor in  stressed power systems. Previous studies have proposed control strategies for GFM-VSCs to improve transient stability. These approaches typically rely on suitable current-limiting algorithms, voltage/reactive-power and active-power supplementary control strategies. This paper investigates and compares the effectiveness of three active-power control strategies in GFM-VSCs  to enhance transient stability in power systems with 100 \% converter-based generation:  (i) a wide-area control strategy (TSP-WACS) using the centre of inertia (COI) frequency, (ii) a local transient damping method (TSP-TDM), and (iii) a novel local control strategy (TSP-L) proposed in this work. All strategies were implemented and assessed using short-circuit simulations on Kundur’s two-area test system with 100\% GFM-VSC generators, demonstrating critical clearing time (CCT)  improvement. The TSP-WACS strategy  achieves the best performance but requires a communication infrastructure, while TSP-L strategy  offers a simple, robust alternative using only local measurements.
\end{abstract}


This is an unabridged draft of the following paper (submitted to 24rd Power Systems Computation Conference, Limassol, Cyprus (PSCC'2026), Electric Power Systems Research Journal, special section: VSI:PSCC 2026, under review):
\begin{itemize}

\item R. E. Ávila-Martínez, L. Rouco, J. Renedo, L. Sigrist A. Garcia-Cerrada, "Active-power control strategies in grid-forming power converters to improve transient stability in power systems with 100\% converter-based generation", submitted to the XXIV Power System Computation Conference (PSCC), Limassol, Cyprus, pp. 1-10, 2026.
\item ID: EPSR-D-25-06392-R1.
\item Internal reference: IIT-25-331WP.
\end{itemize}


\newpage

\section{Introduction}
{Voltage Source Converters with grid-forming control (\ac{GFVSC}) play a crucial role in the stability of power systems with large amounts of converter-based generation~\cite{Paolone2020}. Traditionally, non-synchronous generation resources operate in grid-following (GFL) mode. In contrast, \acp{GFVSC} not only possess the capability for islanded operation but
can also seamlessly connect to the main grid, contributing to the grid strength. Consequently, they must emulate the behavior of a real synchronous generator through by means of appropriate GFM-VSC self-synchronization strategies.} Transient stability (angle stability under large disturbances)  has traditionally been associated with the rotor-angle dynamics of synchronous machines~\cite{Kundur2004,Nikos2021}. In power systems consisting solely of synchronous generators, this phenomenon is well understood and has been  extensively studied. However, as the use of converter-based  resources continues to increase, the concept of transient stability has gained renewed importance in power systems with high proportion of or even 100 \% converter-based generation.  Recent studies have shown that loss of synchronism may also occur in power  systems dominated by grid-forming converters after large disturbances~\cite{YWangTS2020,HAmanoRAS2018}: {if a GFM-VSC emulates a synchronous machine, it could also lose synchronism.} Consequently, the term transient stability is now often used to describe loss-of-synchronism phenomenon in power systems with high or even 100\% penetration of non-synchronous generation~\cite{LasseterRGFMIsTS2019,ZhaoXTSE2020,HeXPanSTS2022} and several publications have examined the impact of various control approaches and parameters on transient stability in \ac{GFVSC}-based systems~\cite{Cheng2020, Eskandari2021}. During the past years, several dedicated control methods in \acp{GFVSC} have been proposed to improve transient stability, which can be classified in the following categories:
\begin{enumerate}
	\item GFM self-synchronization strategies: GFM strategies have an influence on transient stability~\cite{Pan2020,ravila_2026a} and also their parameters (i.e., emulated inertia constant, damping coefficient, etc...). For example, the work in~\cite{Qoria2020} proposed an increase of the emulated inertia constant when a fault occurs to prevent loss of synchronism of the \ac{GFVSC}. Meanwhile, some studies proposed the limitation of the output frequency of the \ac{GFVSC}~\cite{FlynnDZhaoXFreezingGFM2020,FlynnDGFMTS2022,ravila_2026a}, and the work in~\cite{VattaKkuniK2024} proposed the concept of Active-Power Control (VAPC), which significantly improves transient-stability margin of \acp{GFVSC}.
	\item Specific current limiters in the GFM-VSCs: Previous work has shown that conventional vector-control current limiters (Current Saturation Algorithms, CSA-CL) jeopardize transient stability and several current limiters in \acp{GFVSC} have been proposed to improve transient stability~\cite{Qoria_VSC_current_limit2020,Qoria_VSC_CCT2020,Rokrok_TS2022a,Qoria2023, Yang2025}. For example, reference~\cite{Qoria_VSC_current_limit2020} proposed a hybrid current limiter (CSA + virtual-impedance-based current limiter) in \acp{GFVSC} to improve transient stability.  
	\item Supplementary active-power (P) control strategies in GFM-VSCs to improve transient stability~\cite{Shuai2019,Collados-Rodriguez_2023,Yu_2024,Choopani2020,Xiong2021}.
	\item Supplementary reactive-power (Q) / voltage control strategies in GFM-VSCs to improve transient stability~\cite{XiongX_2021a,BlaabjergFTSAngle2022,RAvilaM2022a,SiW2023,RAvilaM2024}. 
\end{enumerate}

In general, methods within the different categories above are complementary and they can be combined. This work just focuses on supplementary active-power (P) control strategies in the GFM-VSCs to improve transient stability. Traditionally, fast reduction of the mechanical active-power of synchronous machines was a very attractive alternative to improve transient stability from a theoretical point of view. However, the practical implementation was very limited, since it is only feasible in steam turbines with fast valving \cite{Kundur1994}. Since \acp{GFVSC} emulate synchronous machines with a control algorithm and the active-power setpoint plays the role of the emulated mechanical active power, and control of power converters can be fast, active-power control strategies in \acp{GFVSC} have an enormous potential to improve transient stability and they are feasible from a practical point of view.  

In the local active-power control strategy proposed in~\cite{Shuai2019}, when the GFM-VSC detects a fault, it reduces its active-power setpoint proportionally to the voltage sag, to prevent loss of synchronism. The work in~\cite{Collados-Rodriguez_2023} proposed an adaptive Active-power/frequency (P-f) droop constant to prevent loss of synchronism of the GFM-VSC. Ref. \cite{Yu_2024} proposed a supplementary P setpoint in the GFM-VSC to improve transient stability, using a proportional-integral-differential (PID) controller with the local frequency deviation as input signal and with an instability-detection method. The work~\cite{Choopani2020} proposes a P-control strategy in GFM-VSCs to improve transient stability, using as input signal the difference between the the frequency of the centre of inertia (COI) and the output frequency of the GFM-VSC and using a PID controller. The proposed P-control strategy uses a Wide Area Control System (WACS) to obtain the COI frequency. This control strategy will be referred as TSP-WACS, for short. \footnote{Transient-stability-tailored active-power control strategies in GFM-VSCs will be referred as \emph{TSP} in this paper.} The authors analyzed the impact of communication latency on the proposed strategy in~\cite{Choopani2019502}. However, this global measurement-based approach presents implementation challenges due to the required communication system.  In contrast, the research presented in~\cite{Xiong2021} proposes a supplementary P-control strategy to improve transient stability using a transient damping method (TDM) for a \ac{GFVSC} connected to an infinite grid (TSP-TDM strategy, for short). The P-control method adds a proportional gain to the frequency deviation with respect to the nominal frequency, using local measurements. The TSP-TDM  strategy changes its active-power set-point proportionally to its frequency deviation through a wash-out filter  as a solution for improving transient stability.  Unlike the strategy in~\cite{Choopani2019502}, which uses global measurements of the centre of inertia to try to pull all GFM-VSC  frequencies together to to converge to  the COI frequency, if a fault produces the increase of the frequencies of the GFM-VSCs,  the TSP-TDM  control action slows down all the converters independently when they are close to or far from the faults. This means that the TSP-TDM  control strategy also slows down the converters with frequency the COI frequency,  jeopardising the transient stability in case of faults far from the GFM-VSC. 

This analysis highlights a noticeable lack of an effective-but-practical  local active power control strategy for GFM-VSCs to improve transient stability
 in multi-converter systems. While the TSP-TDM  controller proposed in~\cite{Xiong2021} provides a local approach, its application and effectiveness in multi-converter systems remain unproven and are still questionable due to its lack of selectivity regarding fault location. Therefore, there is a need to explore effective P-control strategies in \ac{GFVSC}s to improve transient stability in in power systems with 100\% converter-based generation,  especially when using local measurements.

In this context, this paper proposes a local active-power control strategy (TSP-L), designed to be active only for severe-enough faults in a multi-converter system with 100~\% GFM-VSC-based generators. The TSP-L control will be only activated in  those converters that are close to the faults and disabled in  those ones far from the faults. The aim of this design is to ensure a robust performance in terms of transient stability of multi-converter systems in all scenarios. 

Furthermore, this work presents a comparative study of three active-power control strategies in GFM-VSCs  for improving transient stability in power systems with $100\%$ converter-based generation:  the wide-area $\text{TSP-WACS}$~\cite{Choopani2020} (which uses the Centre of Inertia frequency), the local $\text{TSP-TDM}$~\cite{Xiong2021} (based on local frequency deviations), and the novel local $\text{TSP-L}$ (which activates control actions based on local voltage and frequency thresholds).

By conducting a thorough analysis and comparison of various P-supplementary controllers designed for \ac{GFVSC}s, the main contributions of this paper  are as follows:
\begin{itemize}
    \item Proposal of a local active-power control strategy to improve transient stability of power systems with 100\% of \ac{GFVSC}-based generation (TSP-L). 
        \item Comparison of the proposed control strategy with two existing active-power control  strategies based on both global measurements (TSP-WACS) and local measurements (TSP-TDM).
    \item Demonstration, via simulations, of the effectiveness of the control approaches in significantly improving the critical clearing times of different faults.
    \item Analysis of the impact of communication latency on the global control strategy (TSP-WACS), showing that this  strategy is robust for realistic communication delays.
\end{itemize}
The performance of the control strategies is demonstrated through electromagnetic transient simulation (EMT)   using Kundur's two-area test system~\cite{Kundur1994} with 100\% grid-forming VSC-based generation.

\FloatBarrier
\newpage
\section{Grid-forming VSCs}\label{sec:VSC_V}
\subsection{Modelling and control}\label{sec:VSC_V_model}
\begin{figure}[!htbp]
\begin{center}
\includegraphics[width=0.8\columnwidth]{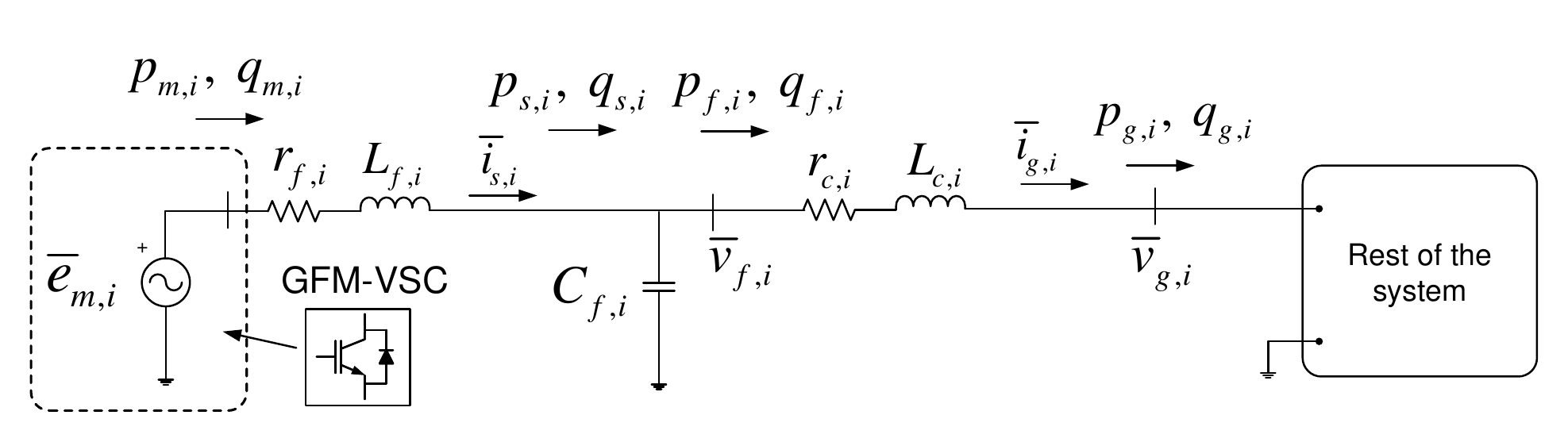}
\caption{Model of a grid-forming VSC. }
\label{fig:VSC_V_model}
\end{center}
\end{figure}

\noindent The equivalent circuit model for a \ac{GFVSC}, following the established methodology in
~\cite{Paolone2020,Qoria2020,Pereira2020}, is described in this section. As depicted in Fig.~\ref{fig:VSC_V_model}, the \ac{GFVSC}-$i$ converter is modeled as a voltage source~($\bar{e}_{m,i}$) connected to the rest of the system through an LC filter, consisting of a phase reactor ($\bar{z}_{f,i}=r_{f,i} + j \omega_{i} L_{f,i}$) and a capacitor ($\bar{z}_{Cf,i}=-j/(\omega_{i} C_{f,i})$), and a transformer ($\bar{z}_{c,i}=r_{c,i} + j \omega_{i} L_{c,i}$).

\begin{figure}[!htbp]
\begin{center}
\includegraphics[width=1\columnwidth]{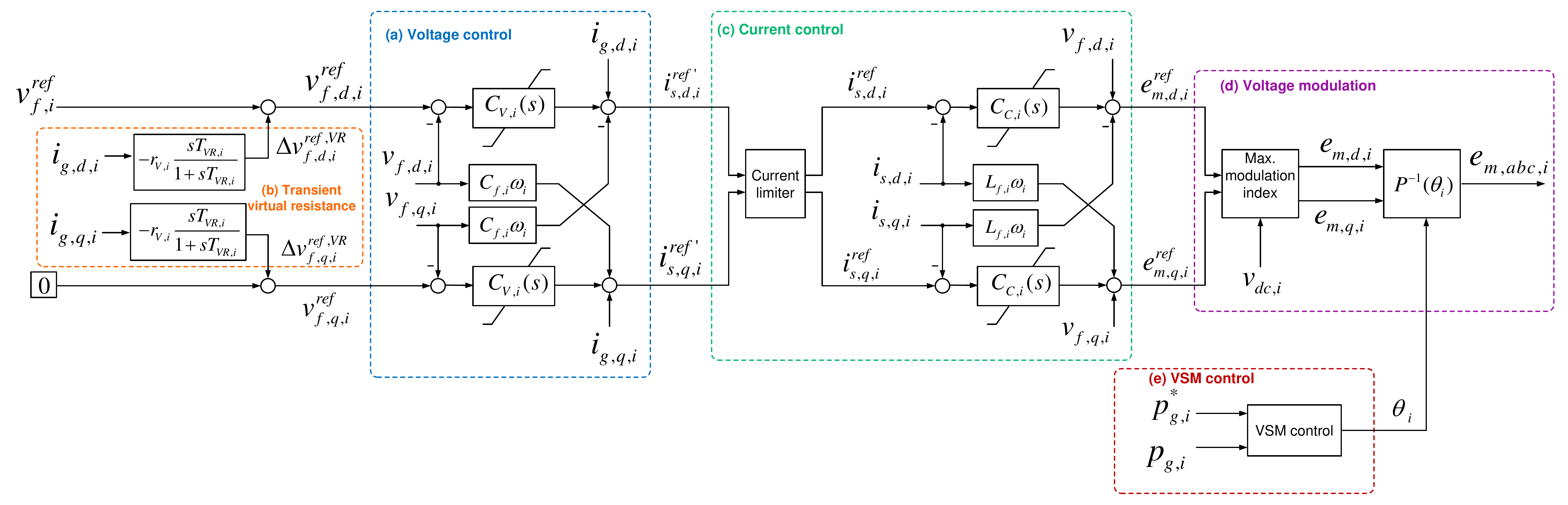}
\caption{General scheme of the control system of a grid-forming VSC. }
\label{fig:VSC_V_control_loops_general}
\end{center}
\end{figure}

Fig.~\ref{fig:VSC_V_control_loops_general} displays the general control scheme for \ac{GFVSC}-$i$,~\cite{Qoria2020}. The scheme consist of a cascade AC voltage and current control loops operating in the $d-q$ reference frame: (a) a voltage controller, (b) a virtual transient resistance to enhance damping during disturbances~\cite{Qoria2020}, (c) a current controller with current limitation, and (d) voltage modulation with modulation index limitations. Finally, (e) a Virtual Synchronous Machine (VSM) control mechanism (\ac{GFVSC} self-synchronization strategy)~\cite{Paolone2020,Qoria2020}.

The \ac{GFVSC}-$i$ controls voltage magnitude $v_{f,i}$ and frequency $\omega_{i}$  (and, thus, the voltage angle $\delta_{f,i}$) at bus $f_{i}$. The angle $\delta_{f,i}$ (rad) is imposed by the \ac{GFVSC} VSM control (e) (Subection~\ref{sec:VSC_V_VSM}), aligning $\bar{v}_{f,i}$ with the direct axis component ($d$-axis) of the rotating $d-q$ reference frame.

\subsection{Virtual synchronous machine control (VSM)}\label{sec:VSC_V_VSM}
\begin{figure}[!htbp]
	\begin{center}
\includegraphics[width=0.9\columnwidth]{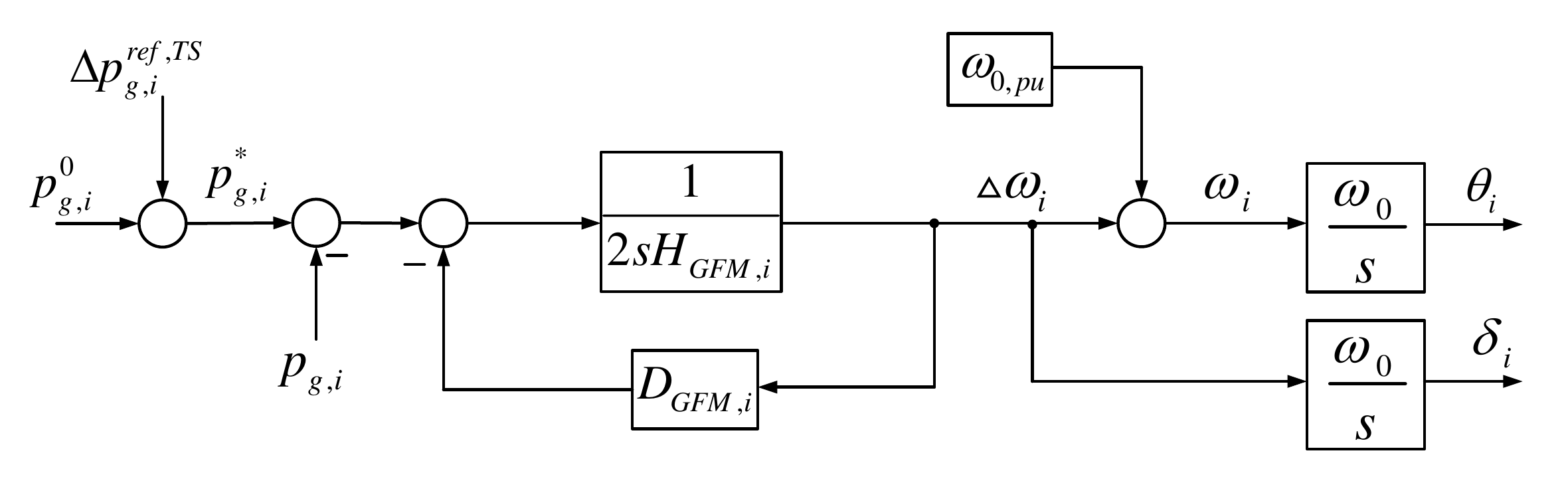}
		\caption{VSM control in a \ac{GFVSC}.}
		\label{fig:VSC_V_VSM}
	\end{center}
\end{figure}

\noindent The virtual synchronous machine (VSM) control is used as a self-synchronization  mechanism in the \ac{GFVSC} as is shown in Fig.~\ref{fig:VSC_V_VSM}~\cite{DArco2015,Choopani2020}. It is the variant \emph{VSM - without PLL}, as used in \cite{Choopani2019502,Xiong2021}.

The swing equation emulated by a VSM is given by:
\begin{eqnarray}\label{eq_VSC_V_VSM_v1}
	2 H_{GFM,i}  \frac{d \Delta \omega_{i}}{dt} & = & p_{g,i}^{*}  - p_{g,i} -D_{GFM,i}  \Delta \omega_{i}
\end{eqnarray}

where:
\begin{itemize}
	\item $H_{GFM,i}$~(s)  is the emulated inertia constant.
	\item $D_{GFM,i}$~(pu)  is the proportional gain of the primary frequency response (PFR). It also plays the role of damping coefficient.
	\item $\Delta \omega_{i}=\omega_{i}-\omega_{0,pu}$~(pu), where $\omega_{i}$ is the frequency imposed by the \ac{GFVSC} and $\omega_{0,pu}=1$~pu.
    \item $p_{g,i}^{*}$~(pu) is the  active-power set-point of the \ac{GFVSC}.
    \item $\Delta p_{g,i}^{ref,TS}$~(pu) is a supplementary active-power set-point for transient stability improvement, which will be discussed in Section~\ref{sec:apcs_VSC_V_TS}. 
	\item $p_{g,i}$~(pu) is the active-power injected by the \ac{GFVSC} at the connection point.
	\item $\omega_0$ is the nominal frequency in rad/s.
\end{itemize}

Through equation (\ref{eq_VSC_V_VSM_v1}), the \ac{GFVSC}-$i$ controls the frequency $\omega_{i}$~(pu) at its connection point (bus $f_{i}$). Fig.~\ref{fig:VSC_V_VSM} shows the angles used for the Park's Transformation applied in the control loops (see Fig.~\ref{fig:VSC_V_control_loops_general}).

\section{Active-Power control strategies}\label{sec:apcs_VSC_V_TS}
It is assumed that the GFM-VSCs are connected to a transmission system (inductive network).  The active-power injection of a certain GFM-VSC-$i$ is given by: 
\begin{equation}\label{eq:VSC_pg_electrical}
	p_{g,i} \simeq \frac{v_{f,i} v_{g,i}}{x_{c,i}} \sin (\delta_{f,i}-\delta_{g,i})
\end{equation}
\noindent where $v_{f,i}$ and $v_{g,i}$ are the voltage magnitudes at  bus $f_i$ and the PCC ($g_i$), respectively; $x_{c,i}$ is the coupling reactance; and $\delta_{f,i}$ and $\delta_{g,i}$ are the corresponding power angles of $v_{f,i}$ and $v_{g,i}$.

During a short-circuit  event, the voltage at the connection point ($v_{g,i}$) decreases, leading to a reduction in active power injection ($p_{g,i}$) according to Eq.~(\ref{eq:VSC_pg_electrical}). This implies that the frequency of GFM-VSC-$i$ increases, according to (\ref{eq_VSC_V_VSM_v1}). 

Notice that, analogously to synchronous machines, in GFM-VSCs the active-power setpoint $p_{g,i}^*$ in Fig. \ref{fig:VSC_V_VSM} and (\ref{eq_VSC_V_VSM_v1}) plays the role of the \emph{emulated mechanical active power}. Therefore, P setpoint $p_{g,i}^*$ could be used to improve transient stability. If $p_{g,i}^*$ decreases (increases),  the frequency of GFM-VSC-$i$ and nearby GFM-VSCs are slowed down (or accelerated, respectively).

The total active-power setpoint of the GFM-VSC,
$p_{g,i}^{*}$ in Fig. \ref{fig:VSC_V_VSM}, is given by:  
\begin{eqnarray}\label{eq:VSC_pg_ref_TS}
p_{g,i}^{*}  = p_{g,i}^{0} + \Delta p_{g,i}^{ref,TS}
\end{eqnarray}

\noindent where:
\begin{itemize}
    \item $p_{g,i}^{0}$ is a constant active-power setpoint. 
    \item $\Delta p_{g,i}^{ref,TS}$ is the supplementary active-power setpoint  for transient stability improvement. 
\end{itemize}

Three active-power control strategies in \acp{GFVSC} to improve transient stability will be analyzed in this work:  
\begin{itemize}
    \item Active-power control using a wide-area control system (\textbf{TSP-WACS})~\cite{Choopani2020} (global measurements). 
    \item Active-power control using transient damping method (\textbf{TSP-TDM})~\cite{Xiong2021} (local measurements). 
    \item A novel active-power control (\textbf{TSP-L}) strategy proposed in this work (local measurements). 
\end{itemize}

\FloatBarrier
\subsection{TSP-WACS strategy~\cite{Choopani2020} }\label{sec:apcs_VSC_V_TS_APC_WACS}

\begin{figure}[!htbp]
\begin{center}
\includegraphics[width=1.0\columnwidth]{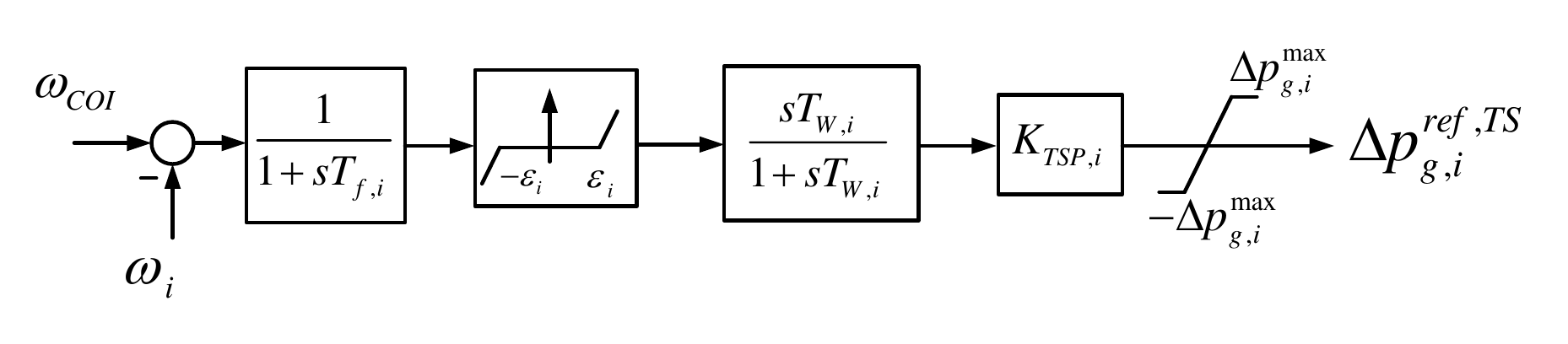}
\caption{Strategy TSP-WACS. }
\label{fig:apcs_VSC_GF_TSP_WACS}
\end{center}
\end{figure}

Fig.~\ref{fig:apcs_VSC_GF_TSP_WACS} shows the TSP-WACS strategy proposed in~\cite{Choopani2020} to improve transient stability. As this strategy involves calculating and using the frequency of the center of inertia (COI), $\omega_{COI}$, frequency, {a} WACS  is required. {In a power system with 100~\% GFM-based generation, the COI frequency is defined as~\cite{Choopani2020}:}
\begin{equation}\label{eq:w_COI}
\omega_{COI}=\frac{1}{H_{tot}}\sum_{k=1}^{n_G} H_{GFM,k} \omega_{k} \mbox{ (pu), with } H_{tot}= \sum_{k=1}^{n_G} H_{GFM,k}.
\end{equation}
The use of the COI frequency has been very useful in WACS-based supplementary controllers to improve transient stability in different contexts, such as excitation boosters in synchronous machines \cite{LuisDM2017, LuisDM2019} or in high voltage direct current systems based on voltage source converters (VSC-HVDC) \cite{Eriksson2014a}, for example. 

Although in~\cite{Choopani2020} the authors proposed a general scheme of TSP-WACS strategy using a PID controller, this paper will use a proportional controller (gain $K_{TSP,i}$), since it proved to be remarkably effective. In addition, this work includes blocks that are often useful in transient-stability controllers: a low-pass filter (with time constant $T_{f,i}$), a wash-out filter (with time constant $T_{W,i}$), a deadband ($\pm \epsilon_i$), and a saturator ($\pm p_{g,i}^{max}$).  

The objective of TSP-WACS strategy  is to manipulate the active-power setpoint of each GFM-VSC to pull the frequencies of all converters towards the COI frequency, thereby preventing loss of synchronism among them. The control action of Fig.~\ref{fig:apcs_VSC_GF_TSP_WACS} ($\Delta p_{g,i}^{ref,TS}$) is determined based on the deviation of the local frequency ($\omega_{i}$) of GFM-VSC-$i$ with respect to  the COI frequency ($\omega_{COI}$). According to the emulated swing equation of the synchronization strategy (Fig. \ref{fig:VSC_V_VSM} and (\ref{eq_VSC_V_VSM_v1})), by reducing (increasing) the active-power setpoint of each GFM-VSC-$i$, its frequency slowed down (or accelerated).  

The behaviour of TSP-WACS strategy can be summarized as follows:  
\begin{itemize}
    \item If the fault (short circuit) is close to GFM-VSC-$i$, the voltage sag is more severe, and the frequency of the GFM-VSC-$i$ increases more,  resulting in a frequency above the COI frequency ($\omega_{i} > \omega_{COI}$). Then, TSP-WACS strategy adds a negative supplementary active-power setpoint ($\Delta p_{g,i}^{ref,TS} < 0$) to slow down the frequency GFM-VSC-$i$. 
    \item If the fault (short circuit) is far from GFM-VSC-$i$, the voltage sag is smaller, and the frequency of the GFM-VSC-$i$ increases less than in other converters of the system,  resulting in a frequency below the COI frequency ($\omega_{i} < \omega_{COI}$). Then, TSP-WACS strategy adds a positive supplementary active-power setpoint ($\Delta p_{g,i}^{ref,TS} > 0$) to accelerate the frequency of the GFM-VSC-$i$. 
    \item In this way, control actions will pull together the frequencies of \ac{GFVSC}s of the system.
\end{itemize}

\subsection{TSP-TDM strategy~\cite{Xiong2021} }\label{sec:apcs_VSC_V_TS_APC_TDM}
\begin{figure}[!htbp]
\begin{center}
\includegraphics[width=1.0\columnwidth]{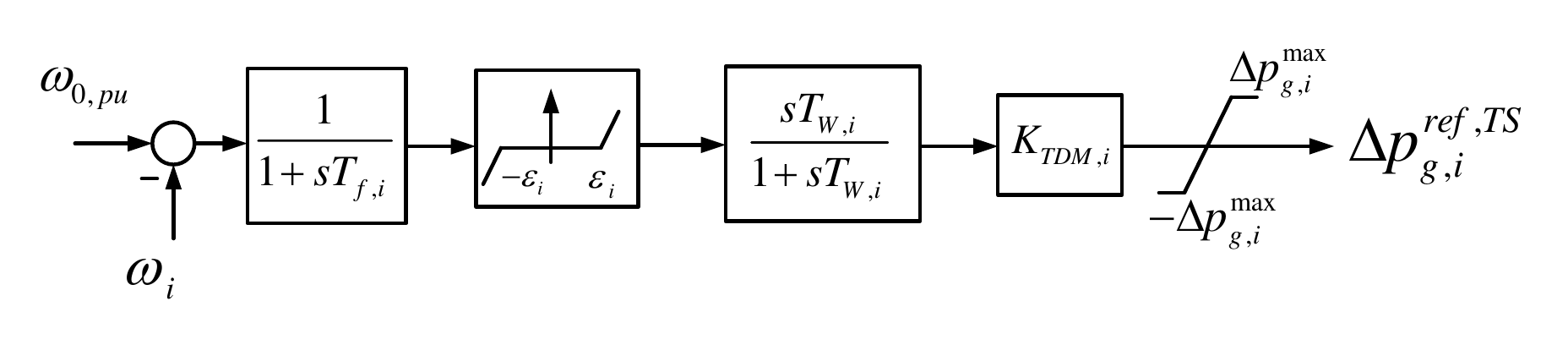}
\caption{Strategy TSP-TDM. }
\label{fig:apcs_VSC_GF_TSP_TDM}
\end{center}
\end{figure}

This control strategy consists of a transient damping method in VSM proposed in~\cite{Xiong2021} for transient stability improvement in GFM-VSCs connected to infinite grid.  The input of the controller of TSP-TDM is the frequency deviation of each \ac{GFVSC}-$i$, with respect to an absolute reference frequency, $\Delta\omega_i = \omega_i - \omega_{0,pu}$. Fig.~\ref{fig:apcs_VSC_GF_TSP_TDM} shows the scheme of the TDM implemented in a \ac{GFVSC} with VSM control.

The TSP-TDM  control strategy consists of a gain ($K_{TDM,i}$)  and a wash-out filter (with a time constant of $T_{W,i}$). Although not discussed in \cite{Xiong2021}, this work includes a saturation parameter to limit the contribution of control actions, denoted as $\pm \Delta p_{g,i}^{max}$, and a deadband ($\pm \epsilon_i$). A low-pass filter (with time constant $T_{f,i}$) is also considered, although the authors do not use it ($T_{f,i}=0$ s), in order to produce a pure damping effect.  The controller takes the frequency deviation ($\Delta \omega_i$) calculated by the VSM algorithm of the \ac{GFVSC} as input, and its output is a supplementary active-power set point ($\Delta p_{g,i}^{ref,TS}$). In the TSP-TDM   control strategy, each \ac{GFVSC} adds a supplementary active-power set point that is proportional to the frequency deviation, which uses only local information. This means that, if a short-circuit occurs, the frequency of the GFM-VSC $\omega_i$ increases and TSP-TDM strategy adds a supplementary P setpoint $\Delta p_{g,i}^{ref,TS} < 0$, slowing down the frequency of the GFM-VSC. 

TSP-TDM strategy improves transient stability (large disturbance), but it also produces a  pure damping effect on the dynamic response of the GFM-VSC (small-disturbance stability).   However,  in contrast to the TSP-WACS, which uses global measurements of the centre of inertia to try to pull all GFM-VSC   frequencies together to join the COI frequency,  if a fault increases the frequencies of the system, the TSP-TDM  control action would slow down the frequencies of  all the converters independently when they are close to or far from the faults. This means that the TSP-TDM  control strategy also slows down the converters below the centre of inertia, which would jeopardize the transient stability in some cases. 

\newpage

\subsection{Proposed  local active-power control  (TSP-L)}\label{sec:apcs_VSC_V_TS_APC_L}
\begin{figure}[!htbp]
\begin{center}
\includegraphics[width=0.8\columnwidth]{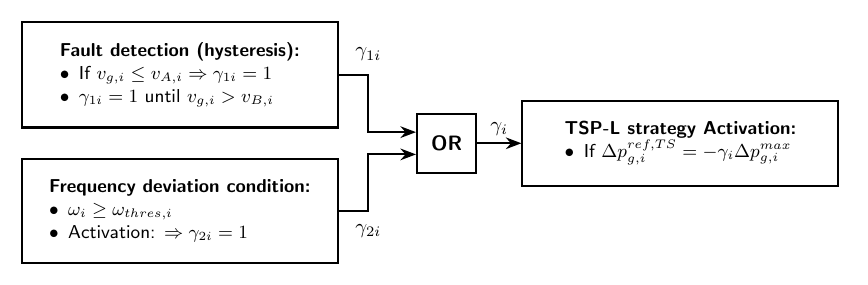}
\caption{Activation philosophy for the TSP-L strategy.}
\label{fig:activation_logic_v2}
\end{center}
\end{figure}

\begin{figure}[!htbp]
\begin{center}
\includegraphics[width=0.9\columnwidth]{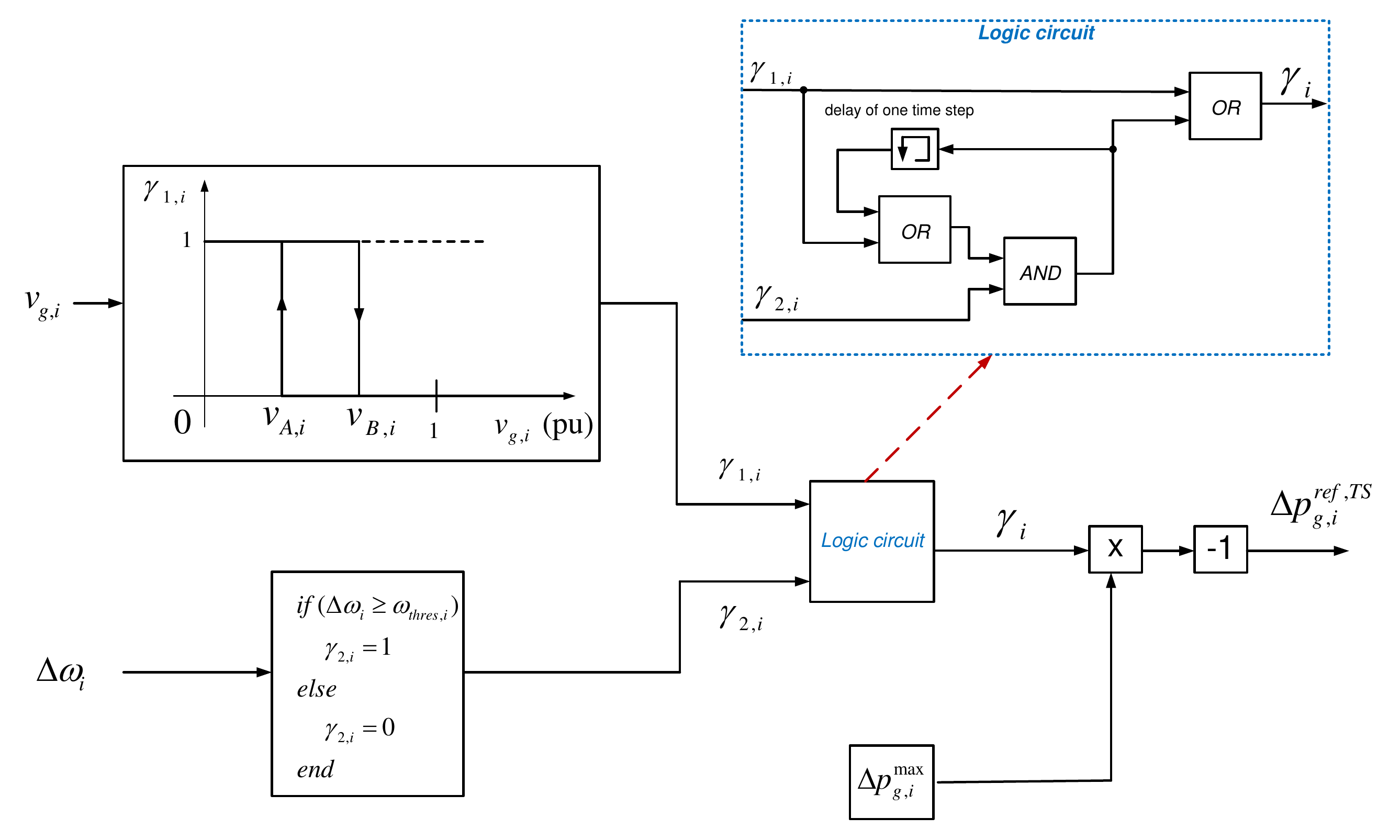}
\setlength{\belowcaptionskip}{-20pt}
\caption{Strategy TSP-L and Logic circuit for fault detection.}
\label{fig:apcs_VSC_GF_TSP_L&Logic}
\end{center}
\end{figure}

This section proposes a  local active-power (TSP-L) control strategy applying a supplementary active-power setpoint, $\Delta p_{g,i}^{ref,TS}$, to the power reference as shown in the flowchart of  Fig.~\ref{fig:activation_logic_v2}. The detailed block diagram of TSP-L strategy is shown in Fig.~\ref{fig:apcs_VSC_GF_TSP_L&Logic}.  TSP-L strategy  uses two input signals: the terminal voltage ($v_{g,i}$) and the frequency deviation ($\Delta \omega_{i}$) of the GFM-VSC-$i$. This combination of inputs is used to create a selective activation mechanism. By monitoring the local voltage and frequency, the TSP-L ensures that the supplementary active power control is only triggered when critical fault thresholds are reached. TSP-L is activated when the voltage at the connection point is lower than a certain threshold and it remains activated if the frequency difference is greater than a certain threshold ($v_{A,i}$, $v_{B,i}$ and $\omega_{thres,i}$).

\newpage

The activation philosophy for the TSP-L strategy shown in Figs.~\ref{fig:activation_logic_v2} and \ref{fig:apcs_VSC_GF_TSP_L&Logic}  is as follows:

\begin{itemize}
	\item Binary variable $\gamma_{1,i}$ is set to 1 if a voltage sag is detected with an hysteresis, as shown in Fig.~\ref{fig:apcs_VSC_GF_TSP_L&Logic}. If $v_{g,i}\leq v_{A,i}$, then $\gamma_{1,i}=1$ and remains equal to 1 until $v_{g,i}>v_{B,i}$. if a fault is not detected, then $\gamma_{1,i}=0$.
	\item Binary variable $\gamma_{2,i}$ is set to 1 if the frequency deviation of \ac{GFVSC}-$i$ (with respect to the nominal frequency) is greater than or equal to a certain threshold: $\Delta \omega_{i}\ge \omega_{thres,i}$. Otherwise, $\gamma_{2,i}=0$.
	\item The supplementary controller is activated with binary variable $\gamma_{i}$, which is the result of a logic circuit with $\gamma_{1,i}$ and $\gamma_{2,i}$ as inputs, as shown in Fig.~\ref{fig:apcs_VSC_GF_TSP_L&Logic}.
	\item The supplementary active-power set point is given by: $\Delta p_{g,i}^{ref,TS}= -\gamma_{i} \Delta p_{g,i}^{max} $, where $\Delta p_{g,i}^{max}>0$ is a parameter of the controller ($\gamma_{i}=0$ if the controller is deactivated and $\gamma_{i}=1$ if the controller is activated). 
\end{itemize}

The logic used to activate TSP-L  controller can then be summarised as follows:
\begin{itemize}
	\item The controller will be activated if a voltage sag is detected. Therefore, $\gamma_{1,i}$ will drive the activation of the controller.
	\item Once the controller is activated, the supplementary active-power set-point is maintained if at least one of the two following conditions are satisfied: undervoltage ($\gamma_{1,i}=1$) or frequency greater than or equal to the threshold ($\gamma_{2,i}=1$).
    \item The supplementary P setpoint is negative ($\Delta p_{g,i}^{ref,TS}<0$), to slow down the frequency $\omega_i$ of the GFM-VSC to prevent loss of synchronism. 
\end{itemize}

Unlike TSP-TDM, which activates the control more broadly, the key difference with TSP-L lies in its activation logic, which has conditioned action rules. This ensures that the control is only activated during several-enough faults, guaranteeing that supplementary active power is applied only to the converters that require it within the multi-converter environment.

\FloatBarrier
\section{Results}\label{sec:apcs_Results2_VSC_V_Kundur}
\begin{figure}[!htbp]
	\centering
\includegraphics[width=1\columnwidth]{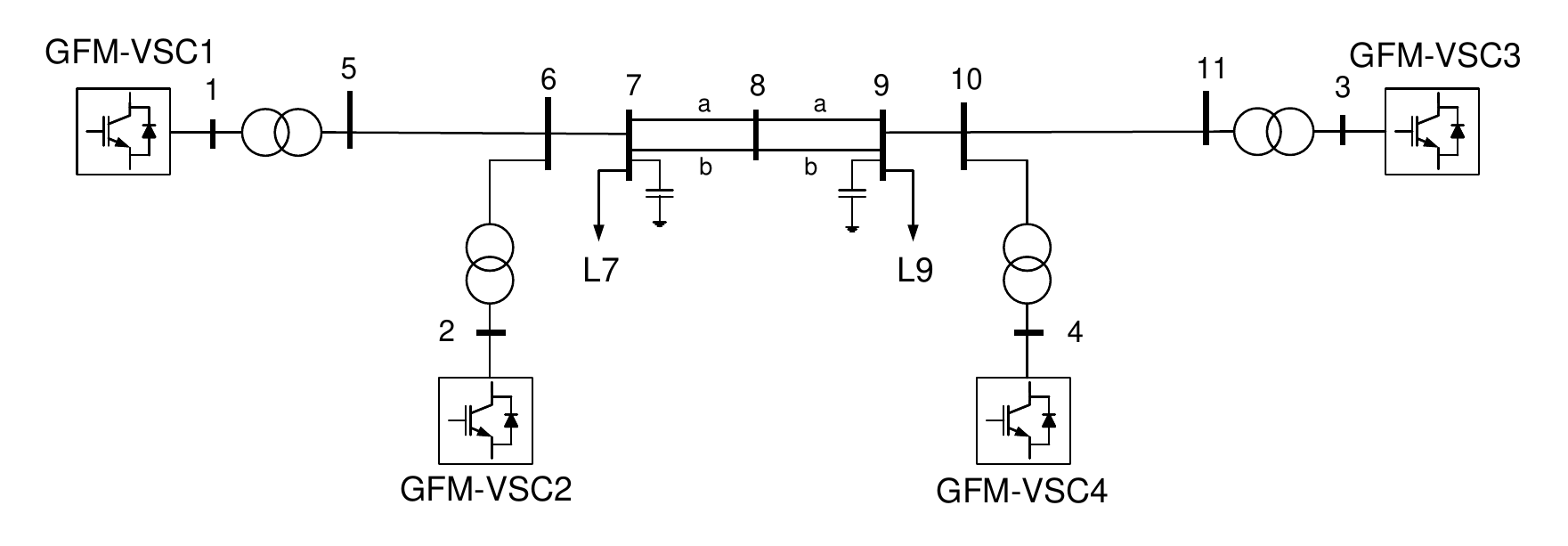}
	\caption{Kundur's two-area test system with 100~\% \ac{GFVSC}-based generation. }
	\label{fig:Kundur_two_area_VSC_V}
\end{figure}

This section assesses the three supplementary $\text{P-control}$ strategies by analyzing the system's transient stability. The analysis is performed on Kundur's two-area test system~\cite{Kundur1994}, where the original synchronous machines have been replaced by 100\% grid-forming VSC-based generation, as illustrated in Fig.~\ref{fig:Kundur_two_area_VSC_V}. The \acp{GFVSC}-based generators use  VSM control and maintain the same nominal apparent power (900 MVA) as the original synchronous generators. System data and \acp{GFVSC} parameters are detailed in the Appendix (Table~\ref{tab:VSC_parameters}). Simulations were conducted using average electromagnetic-type (EMT) models within the VSC\_Lib tool, an open-source library based on $\text{Matlab + Simulink + SimPowerSystems}$~\cite{L2EP_VSC_GF_2020}. 

Four cases are compared:
\begin{itemize}
\item Base case: no supplementary controller for transient stability is implemented in the \ac{GFVSC}s.
\item TSP-L: GFM-VSCs applying TSP-L strategy (Fig.~\ref{fig:apcs_VSC_GF_TSP_L&Logic}, with parameters: $v_{A,i}=0.50$~pu, $v_{B,i}=0.9$~pu, $\omega_{thres,i}=10^{-3}$~pu, $\Delta p_{g,i}^{max}=0.15$~pu.
\item TSP-WACS: GFM-VSCs applying TSP-WACS strategy (Fig.~\ref{fig:apcs_VSC_GF_TSP_WACS}), with parameters: $K_{TSP,i}=100$~pu, $T_{f,i}=0.1$~s, $T_{W,i}=10$~s, $\Delta p_{g,i}^{max}=0.15$~pu and $\epsilon_{i}=10^{-3}$~pu.
\item TSP-TDM: GFM-VSCs applying TSP-TDM strategy (Fig.~\ref{fig:apcs_VSC_GF_TSP_TDM}) with parameters: $K_{TDM,i}=100$~pu,  $T_{f,i}=0$~s,  $T_{W,i}=10$~s, $\Delta p_{g,i}^{max}=0.15$~pu and $\epsilon_{i}=10^{-3}$~pu.
\end{itemize}

Gain values of TSP-WACS and TSP-TDM were tuned to obtain reasonable improvements on transient stability, leading to $K_{TSP}=100$~pu (TSP-WACS) and $K_{TDM}=100$~pu (TSP-TDM). 

\subsection{Short-circuit simulation}\label{sec:apcs_Results2_VSC_V_Kundur_sim1}
\begin{figure}[!htbp]
\centering
\includegraphics[width=0.55\columnwidth]{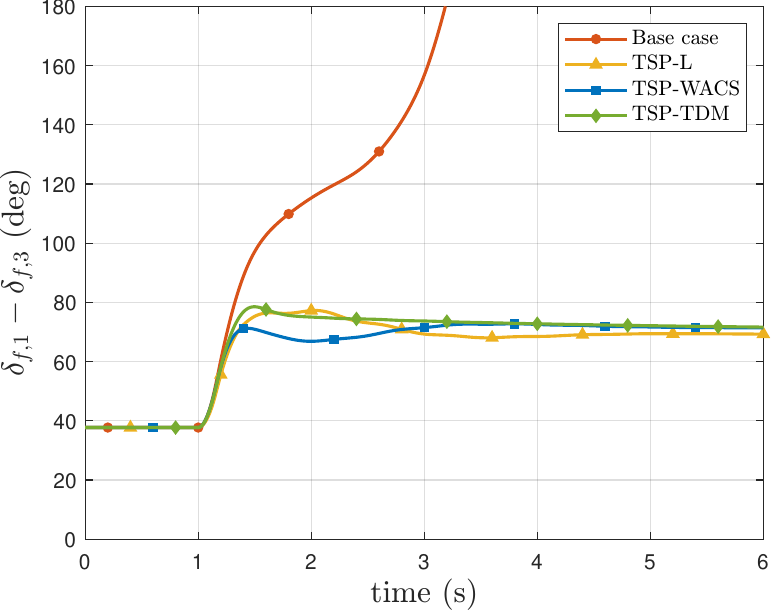}
\caption{Fault I cleared after 150~ms. Angle difference of the GFM-VSCs applying TSP controllers.}
\label{fig:apcs_Kundur_sim1_angles}
\vspace*{0.3cm}
\end{figure}

\noindent Fault I is applied at $t = 1$~s and cleared after $150~\text{ms}$. As shown in Fig.~\ref{fig:apcs_Kundur_sim1_angles} (angle difference between GFM-VSCs 1 \& 3 ), GFM-VSC-based  generators lose synchronism in the Base Case. However, synchronism is successfully maintained when any of the three supplementary controllers ($\text{TSP-L}$, $\text{TSP-TDM}$, and $\text{TSP-WACS}$) are implemented.

\FloatBarrier

{Fig.~\ref{fig:apcs_Kundur_sim1_freq_COI} shows the frequency deviations of the GFM-VSCs  with respect to the $\text{COI}$ frequency. The $\text{TSP-WACS}$ strategy provides the most coordinated action (Fig.~\ref{fig:apcs_Kundur_sim1_deltaPg_Pg_TS}): GFM-VSCs  1 and 2 ($\omega_{i} > \omega_{COI}$) apply  a negative supplementary power set-point to slow down the converters, while GFM-VSCs  3 and 4 ($\omega_{i} < \omega_{COI}$) apply  a positive set-point to accelerate the GFM-VSCs (Fig.~\ref{fig:apcs_Kundur_sim1_freq_COI} and \ref{fig:apcs_Kundur_sim1_deltaPg_Pg_TS}). This coordinated control effectively reduces the angle difference and mitigates the risk of loss of synchronism.

The $\text{TSP-L}$ strategy, illustrated in Fig.~\ref{fig:apcs_Kundur_sim1_deltaPg_Pg_TS}, demonstrates its selective activation: the supplementary active-power set-point ($\Delta p_{g,i}^{ref,TS}<0$)  is only applied  to GFM-VSCs  1 and 2 (close to the fault), and not to GFM-VSCs  3 and 4 (distant from the fault). This is the result of the activation logic (Fig.~\ref{fig:apcs_VSC_GF_TSP_L&Logic}). In contrast, the $\text{TSP-TDM}$ strategy reduces the active-power setpoint  in all converters, using local frequency deviation from nominal frequency ($\omega_{0,pu}=1~\text{pu}$). Since all GFM-VSCs  accelerate during the fault (i.e., $\omega_{i} > \omega_{0,pu}$), the $\text{TSP-TDM}$ action decelerates  all of them simultaneously ($\Delta p_{g,i}^{ref,TS}<0$). As discussed previously, the optimal behaviour, taking TSP-WACS as a reference, would be that GFM-VSCs 3 \& 4 increase their P setpoint, and not reduce it. 

\begin{figure}[!htbp]
\centering
\includegraphics[width=0.5\columnwidth]{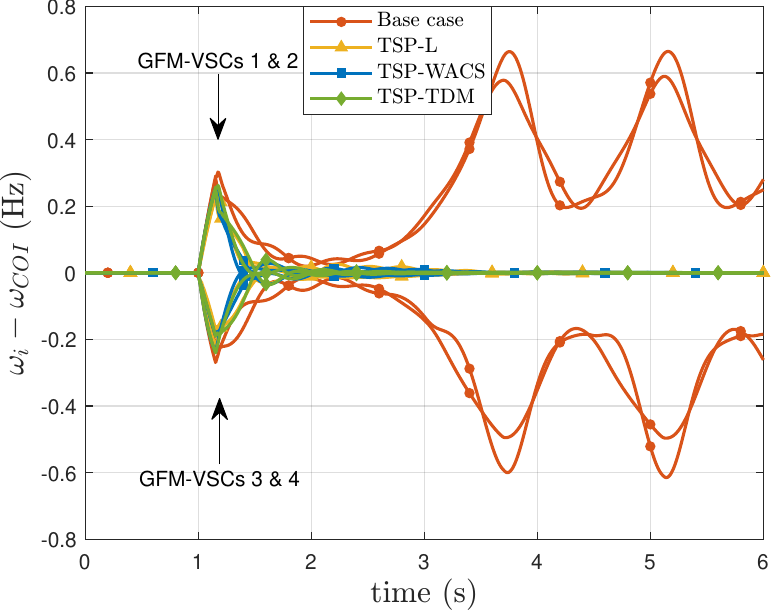}
\caption{Fault I cleared after 150~ms. Frequency deviations of the GFM-VSCs  with respect to the frequency of the COI applying TSP controllers.}
\label{fig:apcs_Kundur_sim1_freq_COI}
\end{figure}
\begin{figure}[!htbp]
\centering
\includegraphics[width=0.9\columnwidth]{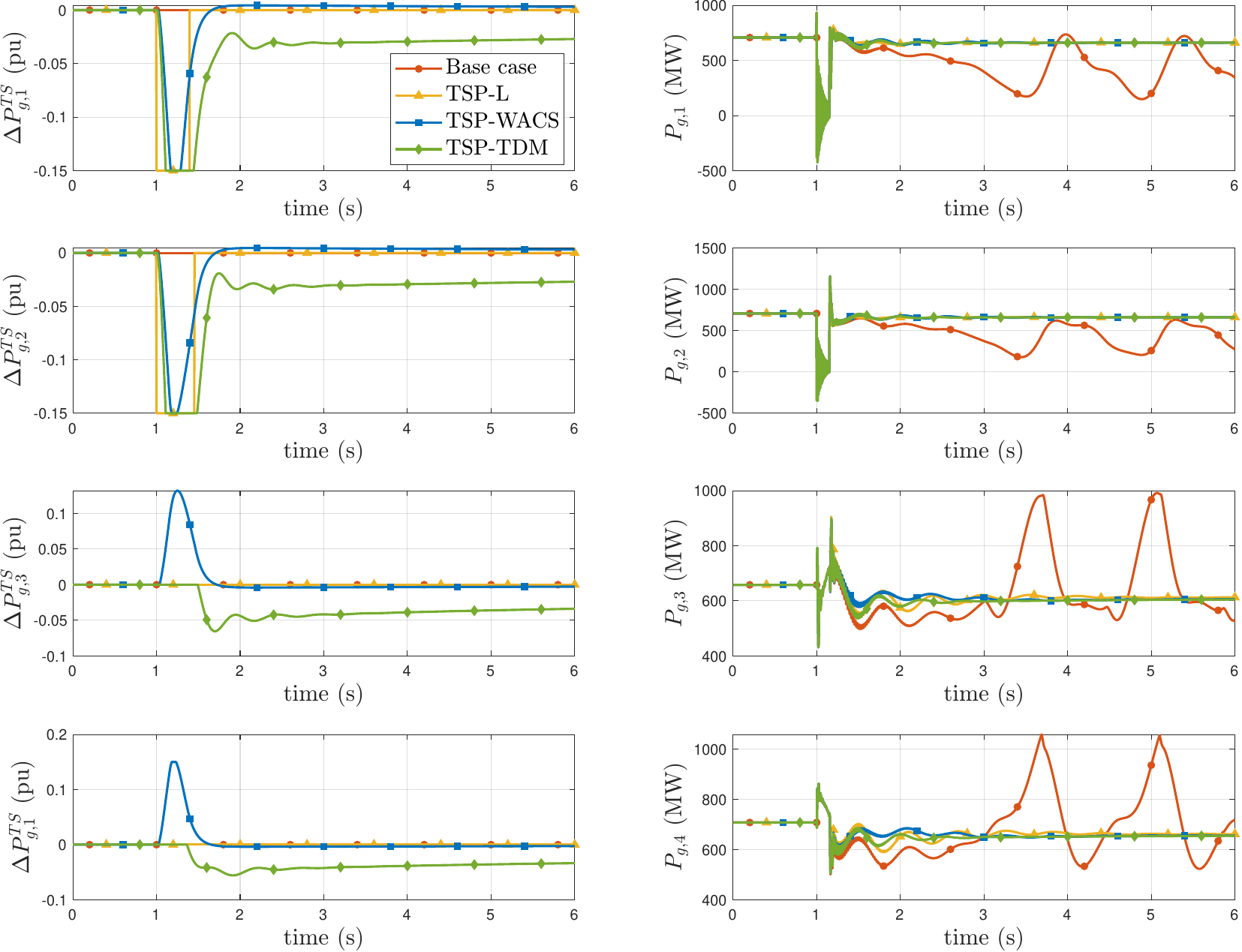}
\caption{Fault I cleared after 150~ms. (left) Active-power set-points and (right) active-power injections of each  GFM-VSC applying TSP controllers.}
\label{fig:apcs_Kundur_sim1_deltaPg_Pg_TS}
\end{figure}

\FloatBarrier
\subsection{Critical clearing times (CCTs)}\label{sec:apcs_Results2_VSC_V_Kundur_CCTs}
\noindent The Critical Clearing Times (CCTs)  for the faults  described in Table~\ref{tab:Kundur_Faults_description} are presented in Table~\ref{tab:apcs_Kundur_CCTs}. All three P-supplementary control strategies ($\text{TSP-L}$, $\text{TSP-TDM}$, and $\text{TSP-WACS}$) significantly increase the $\text{CCTs}$ compared to the base case.  

\begin{table}[!htbp]
	\begin{center}
		\caption{Fault description.}
			\begin{tabular}{|l|c|c|c|}
				\hline
				          & Short circuit & close  & \multirow{2}{*}{clearing  }                                     \\
				          & at line $i-j$ & to bus &                                                \\ \hline
				Fault I   & 7-8a          & 7      & Disconnect 7-8a                                \\
				Fault II  & 5-6           & 5      & short circuit cleared  (line not disconnected) \\
				Fault III & 10-11         & 11     & short circuit cleared  (line not disconnected) \\
				Fault IV  & 8-9a          & 8      & Disconnect 8-9a                                \\
				\hline
			\end{tabular}
    \label{tab:Kundur_Faults_description}
	\end{center}
\end{table}

\begin{table}[!htbp]
\begin{center}
\caption{Critical clearing times (CCTs) for TSPs.}
\begin{tabular}{|l|c|cc|c|ccc|}
\hline
\textbf{CCT}   &  \textbf{base}  & \multicolumn{2}{|c|}{\textbf{TSP-L} } &\multirow{2}{*}{\textbf{TSP-TDM}} &\textbf{TSP-WACS}  &       with & delay        \\
    (ms)              & \textbf{case}  & $v_{A,i}=0.75$~pu          & $v_{A,i}=0.5$~pu           && $\tau=0$ ms    & $50$ ms & $100$ ms  \\ \hline
Fault I     & 130 & 300 & 300 & 230 & 280  & 270  & 270 \\
Fault II    & 270 & 390 & 390 & 350 & 430  & 410  & 390 \\
Fault III   & 220 & 260 & 260 & 240 & 320  & 300  & 270 \\
Fault IV    & 420 & 300 & 420 & 540 & 1520 & 1420 & 1410 \\
\hline
\end{tabular}
\label{tab:apcs_Kundur_CCTs}
\end{center}
\end{table}

As Table~\ref{tab:apcs_Kundur_CCTs} illustrates, the effectiveness of the strategies varies by fault type. For instance, the $\text{TSP-WACS}$ strategy yields the best performance for Fault I and Fault II. The $\text{TSP-TDM}$ strategy also achieves good $\text{CCTs}$, reaching a competitive performance for Fault II. However, the $\text{TSP-L}$ strategy, with its dependency on the voltage activation threshold ($v_{A,i}$), shows the most robust increase in $\text{CCTs}$ for Faults I, II, and III.

The performance for Fault IV highlights a key design difference. The Base Case $\text{CCT}$ for Fault IV is notably lower, and the $\text{TSP-L}$ strategy, when implemented with $v_{A,i}=0.75~\text{pu}$, leads to a reduction in the $\text{CCT}$ (from 420 ms to 300 ms, as shown in the table). This adverse effect is mitigated by setting the threshold to a lower value, $v_{A,i}=0.5~\text{pu}$, which maintains the base case $\text{CCT}$ of 420 ms. This demonstrates the importance of the $\text{TSP-L}$'s selective activation, ensuring that the control action only engages for severe faults near the converter without jeopardizing stability elsewhere. Therefore, $v_{A,i}=0.5~\text{pu}$ is adopted as the optimal setting for the $\text{TSP-L}$ strategy. As a general rule for tuning parameter $v_{A,i}$ of the hysteresis characteristic of TSP-L strategy, it should be small enough in such way that the GFM-VSC only detects short circuits close to the converter and, naturally, the value of this parameter may depend on each particular power system. 

\subsection{Impact of communication latency}\label{sec:apcs_Results2_VSC_V_Kundur_delay}
\noindent It is essential to analyze the influence of communication latency on the performance of the $\text{TSP-WACS}$ strategy, as it relies on a WACS.  To model this, the input error signal of the supplementary controller in Fig.~\ref{fig:apcs_VSC_GF_TSP_WACS} is subjected to a communication delay ($\tau$):
\begin{equation}\label{eq:FB_WACS_w_ref_TS_delay}
u_{i}=e^{-s \tau} (\omega_{COI}-\omega_{i})
\end{equation}
Previous studies on $\text{WACS}$ report delays ranging from $50$ to $80~\text{ms}$~\cite{Chow2015}. Based on this, we analyze the $\text{TSP-WACS}$ strategy using conservative delays of $50~\text{ms}$ and $100~\text{ms}$.

The results, presented in the last two columns of Table~\ref{tab:apcs_Kundur_CCTs},  demonstrate that the $\text{TSP-WACS}$ strategy is highly robust against realistic communication latency. While an increase in $\tau$ naturally leads to a slight reduction in $\text{CCTs}$ across various fault scenarios, the strategy consistently maintains the best overall performance compared to $\text{TSP-L}$ and $\text{TSP-TDM}$ (see Table~\ref{tab:apcs_Kundur_CCTs} for comparison).

\FloatBarrier
\subsection{Impact of the activation of the TSP controllers in different areas.}\label{sec:apcs_Results2_VSC_V_Kundur_Areas}

To evaluate the impact of fault location and control selectivity, Tables \ref{tab:apcs_Kundur_CCTs_Area1} and \ref{tab:apcs_Kundur_CCTs_Area2} present the CCTs when the supplementary control strategies are activated only in Area 1 or Area 2, respectively.

\begin{table}[htbp]
\begin{center}
\caption{Critical clearing times (CCTs) for TSPs. Area 1}
\begin{tabular}{|l|c|c|c|c|}
\hline
\textbf{CCT}   &  \textbf{base}  & \textbf{TSP-L} &\textbf{TSP-TDM} &\textbf{TSP-WACS} \\
    (ms)              & \textbf{case}  &  $v_{A,i}=0.5$~pu         & $K_{TDM,i}=100$~pu &  $K_{TSP,i}=100$~pu  \\ \hline
Fault I  & 130 & 300 & 270  & 230 \\
Fault II  & 270 & 390 & 360  & 350  \\
Fault III  & 220 & 220 & 220  & 320  \\
Fault IV  & 420 & 420 & 1080  & 840  \\
\hline
\end{tabular}
\label{tab:apcs_Kundur_CCTs_Area1}
\end{center}
\begin{center}
\caption{Critical clearing times (CCTs) for TSPs. Area 2}
\begin{tabular}{|l|c|c|c|c|}
\hline
\textbf{CCT}   &  \textbf{base}  & \textbf{TSP-L} &\textbf{TSP-TDM} &\textbf{TSP-WACS} \\
    (ms)              & \textbf{case}  & $v_{A,i}=0.5$~pu         & $K_{TDM,i}=100$~pu  & $K_{TSP,i}=100$~pu  \\ \hline
Fault I  & 130  & 130 & 0 & 230  \\
Fault II  & 270  & 270 & 250  & 350  \\
Fault III  & 220  & 260 & 250  & 610  \\
Fault IV  & 420 & 420 & 0 & 790  \\
\hline
\end{tabular}
\label{tab:apcs_Kundur_CCTs_Area2}
\end{center}
\end{table}

In Area 1, all strategies demonstrate a clear improvement over the Base Case for faults I and II. Specifically, the $\text{TSP-TDM}$ and $\text{TSP-WACS}$ strategies achieve substantial $\text{CCT}$ increases for Fault IV, underscoring the potential of active power control when a fault is physically distant from the converters being controlled (Area 1 GFM-VSCs  are far from Fault IV). Conversely, the $\text{TSP-L}$ strategy, designed for selective activation, maintains the Base Case $\text{CCT}$ for Fault IV, confirming that its logic successfully prevents unnecessary control action for less critical disturbances.

The analysis of Area 2 (Table \ref{tab:apcs_Kundur_CCTs_Area2}) reveals a critical drawback of the $\text{TSP-TDM}$ strategy. While $\text{TSP-WACS}$ consistently improves $\text{CCTs}$ across all four fault scenarios, the $\text{TSP-L}$ strategy is correctly inactive for Faults I, II, and IV, maintaining Base Case performance, but successfully increases the $\text{CCT}$ for Fault III. Furthermore,  the $\text{TSP-TDM}$ strategy results in a $\text{CCT}$ of $\mathbf{0~\text{ms}}$ for Faults I and IV. A $\text{CCT}$ of $0~\text{ms}$ indicates that the control action itself destabilizes the system following even a minimal fault clearance time, demonstrating  limitations  of $\text{TSP-TDM}$ strategy  in this multi-converter system.

\FloatBarrier
\subsection{Discussion on the use of the three active-power control strategies}\label{sec:apcs_discussion}
\noindent The choice among the three supplementary active power controllers depends primarily on the trade-off between implementation complexity (cost and infrastructure) and transient-stability  performance ($\text{CCTs}$). The $\text{TSP-L}$ and $\text{TSP-TDM}$ strategies utilize only local measurements, making them easier and more cost-effective to implement as they avoid the need for a Wide-Area Control System ($\text{WACS}$) and its associated communication infrastructure. Conversely, the $\text{TSP-WACS}$ strategy relies on global measurements, adding complexity and cost due to the required communication system, though it provides the best performance.

In terms of effectiveness, the $\text{TSP-WACS}$ strategy, which references the Center of Inertia ($\text{COI}$) frequency, achieves the best results ($\text{CCT}$ improvements). This is because it aligns the control action with the ideal response: applying a negative supplementary active-power setpoint ($\Delta p_{g,i}^{ref,TS}<0$) to slow down the frequency of those GFM-VSCs with  $\omega_{i} > \omega_{COI}$ and applying a positive supplementary active-power setpoint ($\Delta p_{g,i}^{ref,TS}>0$) to accelerate the frequency of those GFM-VSCs with  $\omega_{i} < \omega_{COI}$.  The local strategies, $\text{TSP-L}$ and $\text{TSP-TDM}$, approximate this ideal action using local frequency deviations from nominal frequency. A key advantage of the novel $\text{TSP-L}$ strategy over $\text{TSP-TDM}$ is its selective activation mechanism. By tuning the voltage and frequency thresholds, $\text{TSP-L}$ ensures the controller only activates  for severe, nearby faults by applying a negative supplementary active-power setpoint ($\Delta p_{g,i}^{ref,TS}<0$).  This selectivity prevents the unnecessary intervention observed in the $\text{TSP-TDM}$ strategy, which can adversely affect $\text{CCTs}$ for remote faults.

In conclusion, while the $\text{TSP-WACS}$ strategy is superior in maximizing $\text{CCT}$ and is robust against latency, its deployment is limited by communication requirements. The local $\text{TSP-TDM}$ provides simplicity and clearly improves transient stability when the faults are close and the strategy improves the damping of the system,  but lacks control selectivity. The proposed $\text{TSP-L}$ strategy offers the optimal balance, providing robust $\text{CCT}$ improvements comparable to $\text{TSP-WACS}$ for key fault types, but achieving this without any communication overhead. Therefore, $\text{TSP-L}$ strategy is an effective-but-practical  choice for enhancing transient stability in systems where the implementation of a $\text{WACS}$ is unfeasible.

\FloatBarrier
\section{Conclusions}\label{sec:apcs_conclusion}
\vspace{-0.1cm}
\noindent This paper investigated three supplementary active power control ($\text{P-control}$) strategies in grid-forming power converters  ($\text{TSP-WACS}$, $\text{TSP-TDM}$, and the novel $\text{TSP-L}$) to enhance the transient stability  in power systems with $100\%$ converter-based generation.  A direct comparison of local versus global approaches was provided, including an analysis of $\text{TSP-WACS}$ strategy  sensitivity to communication latency and design guidelines for $\text{TSP-L}$ strategy. 

The key conclusions derived from this paper  are as follows:
\begin{itemize}
    \item The $\text{TSP-WACS}$ strategy, which utilizes global measurements (COI frequency), achieved the best overall performance by effectively coordinating the frequency of all converters, significantly increasing Critical Clearing Times ($\text{CCTs}$), and it proved to be robust against communication latency. 
    \item While the local $\text{TSP-TDM}$ method improved transient stability for close faults, its non-selective action on fault location demonstrated that it could jeopardize transient stability in some cases (remote faults). 
    \item The proposed $\text{TSP-L}$ strategy successfully overcomes the limitations of $\text{TSP-TDM}$ by implementing an activation logic based on local voltage and frequency thresholds. This selectivity ensures the controller acts only for critical faults, establishing $\text{TSP-L}$ as a robust, communication-free alternative for $\text{CCT}$ enhancement.
    \item Design guidelines were provided for parameter tuning TSP-L strategy (parameters of the hysteresis characteristic for fault detection), to ensure its robustness. 
\end{itemize}

\section{Future work  }\label{sec:apcs_future_work}

The work presented
in this paper opens several research lines:
\begin{itemize}
    \item The active-power control strategies in GFM-VSCs to improve transient stability are appropriate when the GFM-VSCs are connected to a transmission system (inductive network). An interesting research line would be control strategies in GFM-VSCs to improve transient stability in resistive AC microgrids.
    \item In this work, an infinite DC-voltage source is used in the GFM-VSCs. An interesting research line would be analysing P-control strategies in GFM-VSC for transient stability taking into account the primary energy / energy storage / DC-voltage control of the GFM-VSCs, studying different technologies.
    \item Further research on active-power control strategies in GFM-VSCs to improve transient stability, including different strategies, other methods (for example, optimization, marching learning, neural networks, etc...), analysis of larger power systems and hardware-in-the loop tests.
\end{itemize}

\section{Acknowledgements}\label{sec:acknowledgements}
Work supported by the Spanish Government under a research project ref. PRE2019-088084, RETOS Project, Spain Ref. RTI2018-098865-B-C31 (MCI/AEI/FEDER, EU) and research project PID2021-125628OB-C21 (MICIU/AEI /10.13039/501100011033 and FEDER, EU); and by Madrid Regional Government under PROMINT-CM Project, Spain Ref. S2018/EMT-4366.

\section*{Appendix: data}
\noindent Table~\ref{tab:VSC_parameters} depicts the data of the GFM-VSCs.  The data of the original two-area Kundur's test system can be found in~\cite{Kundur1994}.
 In this work the same conditions of~\cite{RAvilaM2022a} were considered: (Load at bus 7: 917 MW \& 100 MVAr; load at bus 9: 1817 MW \& 100 MVAr). Nominal frequency is 50 Hz.

\FloatBarrier
\begin{table}[!htbp]
	\caption{Parameters of the GFM-VSCs.}
	\begin{center}
		\renewcommand{\arraystretch}{1.2}
			\begin{tabular}{|l|c|}
				\hline
				\textbf{Parameters} & \textbf{Values} \\
				\hline
				Rating VSC, DC voltage, AC voltage & 900 MVA, $640$ kV, $300$ kV \\
				Max. modulation index ($m_{i}^{max} = \sqrt{\frac{3}{2}} \cdot \frac{V_{dc,B}}{2 V_{ac,B}}$) & 1.31 pu \\
				Series filter resistance ($r_{f,i}$)/reactance ($x_{f,i}$) & 0.005 pu / 0.15 pu \\
				Shunt filter capacitance ($C_{f,i}$) & 0.0660 pu \\
				Transformer resistance ($r_{c,i}$)/reactance ($x_{c,i}$) & 0.005 pu / 0.15 pu \\
				(900 MVA 300/220 kV transformer) & \\
				Current prop./int. control ($K_{C,P,i}$/$K_{C,I,i}$) & 0.73 pu / 1.19 pu/s \\
				Voltage prop./int. control ($K_{V,P,i}$/$K_{V,I,i}$) & 0.52 pu / 1.16 pu/s \\
				Virtual transient resistance ($r_{V,i}$/$T_{VR,i}$) & 0.09 pu / 0.0167 s \\
				Emulated inertia ($H_{GFM,i}$) of GFM-VSCs 1, 2, 3 \& 4  & 4.5 s / 4.5 s / 4.175 s / 6.175 s \\
				Damping constant  ($D_{GFM,i}$) & 20 pu \\
				*GFM-VSC's rating: base values for pu & \\
				\hline
			\end{tabular}
		\label{tab:VSC_parameters}
	\end{center}
\end{table}



\end{document}